# Miniature Organic Transistors With Carbon Nanotubes as Quasi-One Dimensional Electrodes

Pengfei Qi, Ali Javey, Marco Rolandi, Qian Wang, Erhan Yenilmez, and Hongjie Dai*

*Department of Chemistry and Laboratory for Advanced Materials, Stanford University, Stanford, CA 94305, USA*

RECEIVED DATE (automatically inserted by publisher); Email: hdai@stanford.edu

Small organic molecules and conjugated polymers can be easily processed to afford functional electronics such as field effect transistors (FETs),[1a] and in principle, scaling[1b] to single-molecule long devices could circumvent the low carrier mobility problem for these materials to afford high performance ballistic FETs[2,3]. For highly scaled molecular transistors with short channels however, it is crucial to develop novel device geometries to optimize gate electrostatics needed for ON/OFF switching.[4,5]

It is shown here that single-walled carbon nanotubes (SWNT) can be used as quasi one-dimensional (1D) electrodes to construct organic FETs with molecular scale width (~2 nm) and channel length (down to 1-3 nm). The favorable gate electrostatics associated with the sharp 1D electrode geometry allows for room temperature conductance modulation by orders of magnitude for organic transistors that are only several-molecules in length, with switching characteristics superior to devices with lithographically patterned metal electrodes. We suggest that carbon nanotubes may prove to be novel electrodes for a variety of molecular devices.

We first developed a reproducible method of cutting metallic SWNTs to form small gaps within the tubes and with control over the gap size down to $L\sim 2$ nm. The cutting relied on electrical break-down[6] of individual SWNTs between two metal electrodes (Fig. 1a), and the size of the cut was found to be controllable by varying the lengths of the SWNTs (see Ref.6b and Supp. Info). Organic materials were then deposited to bridge the gap in the vapor (for pentacene) or solution phase (for regio-regular ploy (3-hexylthiophene), P3HT), forming the smallest organic FETs with effective channel length down to $L\sim 1$-3 nm and width ~2 nm.

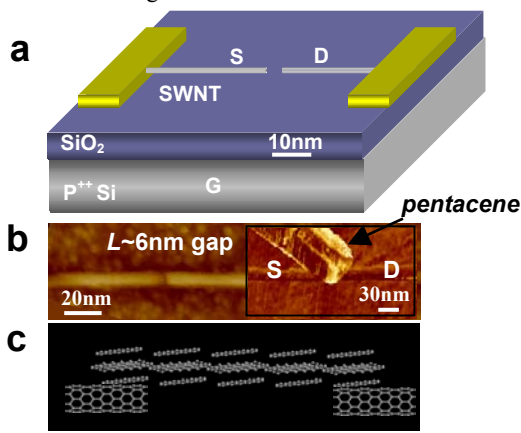

*Figure 1.* SWNTs as electrodes for small organic FETs. (a) Schematic drawing of a cut SWNT with a sub-10 nm gap to be used as source (S) and drain (D) electrodes of an organic FET. The doped Si serves as a back-gate (G) and the SiO$_2$ thickness $t_{ox}$=10 nm. (b) AFM image of a cut SWNT (diameter ~ 2 nm, gap size $L$ measured to be ~ 6 nm after correction of the tip size effect). Right inset: AFM image of a vapor-deposited pentacene crystallite bridging a cut SWNT (see Supp. Info). (c) Drawing of a pentacene crystallite bridging two SWNT electrodes.

With a cut metallic SWNT (gap $L \sim$ 5-6 nm) bridged by a pentacene nano-crystallite (Fig.1b&c), we observed clear semiconducting FET characteristics in the current vs. gate ($I_{ds}$-$V_{gs}$) curve (Fig. 2a). The device exhibited a current modulation of $I_{max}/I_{min} \sim 10^5$ under gating at a fixed bias voltage of $V_{ds}$= -0.5V. The drastic switching clearly differed from the original metallic SWNT device (lack of gate dependence, Fig. 2a inset). This corresponds to the formation of a pentacene FET with channel length $L \sim$ 5-6 nm and width of $w \sim$ 2 nm (i.e., the diameter of the SWNT) as charge transport via hopping between pentacene molecules should be mainly confined in a width on the order of the tube diameter. Notably, the subthreshold swing of the device is S ~ 400 mV/decade (Fig. 2a).

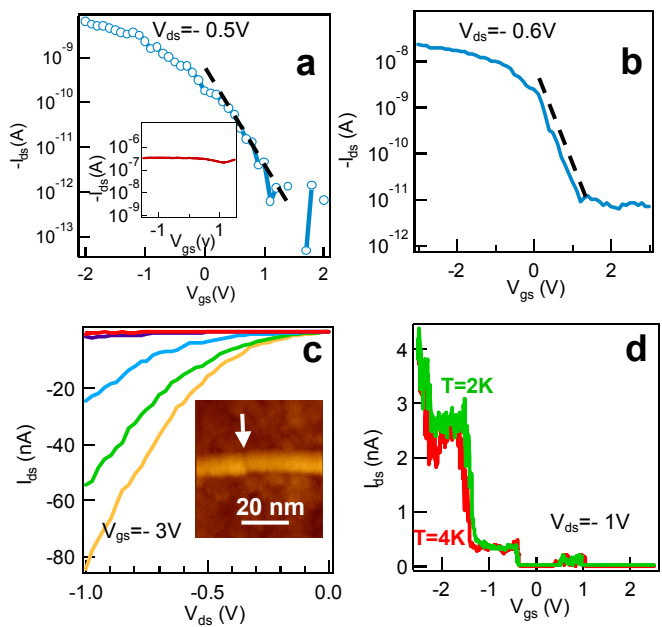

*Figure 2.* Miniature pentacene FETs with metallic SWNTs as source and drain electrodes. (a) Room temperature (T) current ($I_{ds}$) vs. gate-voltage ($V_{gs}$) characteristics of a device (channel length $L\sim$ 6 nm, tube diameter $w \sim$ 2 nm). The dashed line highlights the subthreshold region with S~400mV/decade. Inset: $I_{ds}$-$V_{gs}$ (recorded under bias=10mV) for the metallic SWNT before cutting. (b) Room T $I_{ds}$-$V_{gs}$ characteristics of a $L\sim$ 1-3 nm long pentacene device. S ~ 500 mV/decade indicated by the dashed line. (c) Room T $I_{ds}$-$V_{ds}$ curves for the device in (b) recorded at $V_{gs}$=-3,-2,-1,0 and 1V. Inset: an AFM image of the SWNT after electrical cutting and prior to pentacene deposition. The gap is not fully resolved by AFM. The arrow points to the location of a line-like cut. (d) $I_{ds}$-$V_{gs}$ curves (at $V_{ds}$=1 V) for the same device as in (b)&(c) at T = 4 K and 2 K.

We varied the channel lengths $L$ of SWNT-contacted pentacene FETs ($L \sim$ 1-3 nm, $L \sim$5-6 nm and $L \sim$10-15 nm respectively) and observed length dependent transport properties at various temperatures. At T=300K, the devices exhibited on-current $I_{max}$ scaling approximately with $\sim 1/L$ (under $V_{ds}$=-1 V). This suggests that the active channel lengths for the SWNT-contacted pentacene FETs are set by the gap sizes in the cut tubes and the channel resistance appears to be a dominant part of the resistance of the devices. At T ≤ 4 K, only the shortest pentacene FETs with $L$~1-3 nm consistently (> 5 devices) exhibited appreciable conductance (Fig. 2d), while the longer (L~5-6 nm and L >10 nm) devices became insulating in the entire gate-voltage range. The $L$~1-3 nm devices approach the length of pentacene (~1.2 nm). Transport at low temperature is via tunneling as evidenced by the temperature independent electrical characteristics at T=4 K and 2K (Fig. 2d). Though not fully understood, the step like features in the low temperature $I_{ds}$-$V_{gs}$ data (Fig. 2d) have been observed with multiple samples and could be related to tunneling currents through discrete molecular orbitals of pentacene.

For comparison with SWNT-contacted devices, we fabricated short organic FETs with conventional metal electrodes using electron beam lithography on similar $t_{ox}$~10 nm SiO$_2$/Si substrates. The metal electrodes were ~30-40 nm wide and ~20 nm tall (Pd, with ~1 nm of Ti adhesion layer) with a nano-gap of $L$~sub-10 nm (Fig. 3b inset). Though we were unable to deposit pentacene crystallites to bridge the $L$~sub-10 nm metal gaps due to the poor nucleation of pentacene crystals on the metal electrode, solution dip-coating reliably produced $L$~sub-10 nm long poly-(3-hexylthiophene) (P3HT) FETs with both SWNT and regular metal contacts. SWNT-contacted P3HT FETs exhibited 3 orders of magnitude higher current modulation ($I_{max}/I_{min}$) than the metal contacted devices (Fig. 3a vs. Fig. 3b) over the same $V_{gs}$= –2 V to 2 V gate range. The metal contacted FET can hardly be switched by the gate with a subthreshold swing of S ~ 4 V/decade, while S ~ 400 mV/decade for the SWNT contacted FET under similar bias voltages of $V_{ds}$~0.3 V.

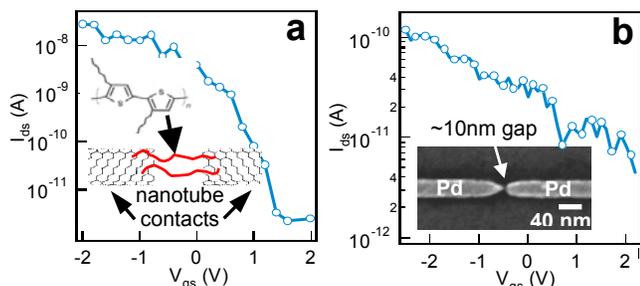

*Figure 3.* Ultra short organic transistors ($L$ ~ 5-10 nm) with SWNT contacts and conventional metal contacts respectively. (a) Room T $I_{ds}$-$V_{gs}$ characteristics for a P3HT FET with SWNT electrodes ($V_{ds}$ =-0.2 V). Diameter of SWNT $w$~ 2 nm and the gap in nanotube $L$ ~ 5-6 nm. S ~ 400mV/decade. (b) $I_{ds}$ vs. $V_{gs}$ characteristics ($V_{ds}$ =-0.4 V) for a P3HT FET with metal (Pd) electrodes with S ~ 4 V/decade. The inset shows an SEM image of the electrodes (width $w$~35 nm, thickness ~ 20 nm, gap size ~ 5-10 nm) used to deposit P3HT and afford the device.

The significantly improved switching characteristics of the short organic FETs with SWNT electrodes over those with metal electrodes are attributed to the excellent electrostatics attainable with the nanotube electrode geometry. It has been shown theoretically that for transistors involving Schottky barriers (SB) at the contacts (which is the case for organic devices), sharp quasi-1D geometry of the source (S) and drain (D) electrodes can facilitate the optimum gate control.[4,5,7,8] Due to the small S/D electrode size, the gate electric field will be able to reach into the channel and regions proximal to the S/D contact junctions to suppress the Schottky barriers (SB) for high ON-states, and to turn OFF the devices efficiently by creating a large body barrier. In contrast, electrostatic gating is ineffective in ultra-short FETs with bulky metal S/D electrodes due to screening of the gate electric fields by the S/D electrodes. Also, the drain potential will be easily penetrating into and dominant over the channel potential, preventing efficient ON and OFF device switching as seen in Fig. 3b.

The electrostatic gating control can be parameterized by a gate efficiency factor α=$C_G$/($C_G$+$C_S$+$C_D$) where $C_G$, $C_S$ and $C_D$ are gate, source and drain capacitances to the channel respectively. The organic devices with SWNT contacts exhibit S~400mV/decade>($k_BT/e$)ln(10)≈60 mV/decade at T=300K, suggesting α is still far from unity and further enhancement in $C_G$ relative to $C_S$ and $C_D$ is needed. The gate insulator in our current devices is un-optimized and thicker than the channel length. It will be feasible to integrate high κ dielectrics such as ZrO$_2$[9] into SWNT-contacted molecular FETs to afford ~10-fold increase in $C_G$ and therefore further enhance the electrostatic gate control. With quasi-1D contacts, gate scaling could allow for organic transistors approaching the limit of S ~ 60 mV/decade.[10]

It is now feasible to fabricate molecular FETs with molecular width to go with the molecular channel length by using SWNTs as electrodes. With quasi-1D SWNT source and drain, scaling could become meaningful to enhance the performance of organic FETs. It is necessary however to form parallel arrays of separated organic devices for high transconductance and currents. Efficient electrostatic gating afforded by quasi 1D electrodes could also facilitate fundamental elucidation of transport in molecules by accessing electronic orbitals within wide energy windows. The intrinsic physical properties of SWNTs could also be utilized for novel contacts. It has been suggested that even with ideal gating (α~1), a limit of S~300mV/decade could exist for molecular transistors with metal (e.g. Au) contacts due to metal induced gap states resulted from metal-molecule interaction.[4] The relatively low density of states of SWNTs may circumvent this problem. Covalent contacts to molecules could also be made via C-C bonding to the nanotube ends. Further, semiconducting SWNTs as electrode contacts could give functions and properties for devices that significantly differ from metallic electrodes. Lastly, it should be possible in the future to develop ultra-small gate electrodes to integrate with SWNT source and drain and achieve various molecular devices with molecular sizes in all dimensions.

**Acknowledgement.** We are grateful to Professor Mark Lundstrom, Jing Guo, Professor Supriyo Datta and Dr. Avik Ghosh for insights. Supported by the MARCO NST Center and an SRC Peter Verhofstadt Graduate Fellowship.

**Supporting Information Available:** Device fabrication, gap formation in SWNTs and bridging gaps by organics. This material is available free of charge via the Internet at http://pubs.acs.org.

**TOC Entry**

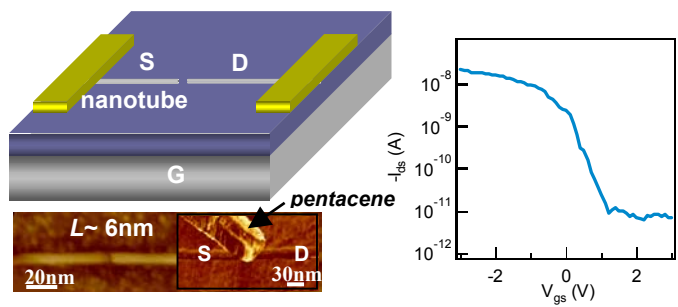

**ABSTRACT FOR WEB PUBLICATION**

As the dimensions of electronic devices approach those of molecules, the size, geometry and chemical composition of the contact electrodes play increasingly dominant roles in device functions.  It is shown here that single-walled carbon nanotubes (SWNT) can be used as quasi one-dimensional (1D) electrodes to construct organic field effect transistors (FET) with molecular scale width (~2 nm) and channel length (1-3 nm). An important feature owing to the quasi 1D electrode geometry is the favorable gate electrostatics that allows for efficient switching of ultra-short organic channels. This affords room temperature conductance modulation by orders of magnitude for organic transistors that are only several-molecules in length, with switching characteristics superior to similar devices with lithographically patterned metal electrodes.  With nanotubes, covalent carbon-carbon bonds could be utilized to form contacts to molecular materials.  The unique geometrical, physical and chemical properties of carbon nanotube electrodes may lead to various interesting molecular devices.